\documentclass[aip,apl,reprint,superscriptaddress]{revtex4-1}
\usepackage{color,units}
\usepackage{amssymb}
\usepackage{amsmath}
\usepackage{psfrag}
\usepackage{topcapt}
\usepackage{hyperref}
\usepackage{float}
\usepackage{braket,color,soul}
\usepackage[normalem]{ulem}

\usepackage{siunitx}

\usepackage[utf8]{inputenc}

\usepackage{ifpdf}

\ifpdf
\usepackage[pdftex]{graphicx}
\else
\usepackage{graphicx}
\fi

\begin{document}
\title{Protecting superconducting qubits from phonon mediated decay}

\author{Yaniv J. Rosen}
\email{rosen10@llnl.gov}
\affiliation{Lawrence Livermore National Laboratory, Livermore, California 94550 USA}

\author{Matthew Horsley}
\affiliation{Lawrence Livermore National Laboratory, Livermore, California 94550 USA}

\author{Sara E. Harrison}
\affiliation{Lawrence Livermore National Laboratory, Livermore, California 94550 USA}

\author{Eric T. Holland}
\affiliation{Lawrence Livermore National Laboratory, Livermore, California 94550 USA}

\author{Allan S. Chang}
\affiliation{Lawrence Livermore National Laboratory, Livermore, California 94550 USA}

\author{Tiziana Bond}
\affiliation{Lawrence Livermore National Laboratory, Livermore, California 94550 USA}

\author{Jonathan L DuBois}
\affiliation{Lawrence Livermore National Laboratory, Livermore, California 94550 USA}

\date{\today}

\ifpdf
\DeclareGraphicsExtensions{.pdf, .jpg, .tif}
\else
\DeclareGraphicsExtensions{.eps, .jpg}
\fi

\begin{abstract}
For quantum computing to become fault tolerant, the underlying quantum bits must be effectively isolated from the noisy environment.  It is well known that including an electromagnetic bandgap around the qubit operating frequency improves coherence for superconducting circuits. However, investigations of bandgaps to other environmental coupling mechanisms remain largely unexplored. Here we present a method to enhance the coherence of superconducting circuits by introducing a phononic bandgap around the device operating frequency. The phononic bandgaps block resonant decay of defect states within the gapped frequency range, removing the electromagnetic coupling to phonons at the gap frequencies. We construct a multi-scale model that derives the decrease in the density of states due to the bandgap and the resulting increase in defect state $T_1$ times. We demonstrate that emission rates from in-plane defect states can be suppressed by up to two orders of magnitude. We combine these simulations with theory for resonators operated in the continuous-wave regime and show that improvements in quality factors are expected by up to the enhancement in defect $T_1$ times. Furthermore, we use full master equation simulation to demonstrate the suppression of qubit energy relaxation even when interacting with 200 defects states. We conclude with an exploration of device implementation including tradeoffs between fabrication complexity and qubit performance. \end{abstract}

\maketitle
Unless explicitly designed to do so, superconducting qubits do not couple directly to phonons. However, nearly all potentially dissipative processes, including interaction of the qubit with lossy dielectrics\cite{Martinis2005,Kim2008} and resistive losses associated with nonequilibrium quasiparticles,\cite{Grunhaupt2018} ultimately rely on phonons to irreversibly carry energy from the system to the environment. Compared to the phonon bath, defect states and unpaired electrons have a relatively sparse density of states. Dissipation canonically occurs when energy is irreversibly transferred to the phonon bath in exchange for an increase of entropy.

A central challenge of qubit fabrication lies in the design of structures and processes that remove or ameliorate uncontrolled coupling to the environment. Examples of efforts to date in superconducting systems include: deep substrate etching,\cite{Bruno2015} undercuts and geometry optimization to reduce participation with lossy dielectrics and surface defects,\cite{Geerlings2012,Sandberg2012} improvements in interface quality,\cite{Megrant2012,Richardson2016} surface processing and vacuum hygiene to reduce the concentration of adsorbates and / or the density of surface and interfacial defects,\cite{Kumar2016} and so on. Despite these efforts, dielectric loss remains a dominant source of relaxation in planar superconducting qubits.\cite{Wenner2011,Wang2015,Dial2016,Sage2011} Furthermore, to date no strategy exists for addressing one of the primary mechanisms of transporting energy to the environment: coupling to the phonon bath.

\begin{figure}
\includegraphics[width=7cm]{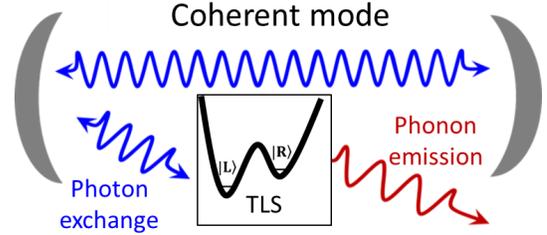}
\caption{Schematic representation of dissipative coupling. A superconducting qubit or resonator can exchange energy with a two-level system defect. This defect couples to phonon modes and can relax to the ground state by emitting a phonon.}
{\label{fig:intro}}
\end{figure}

\begin{figure*}
\includegraphics[width=17cm]{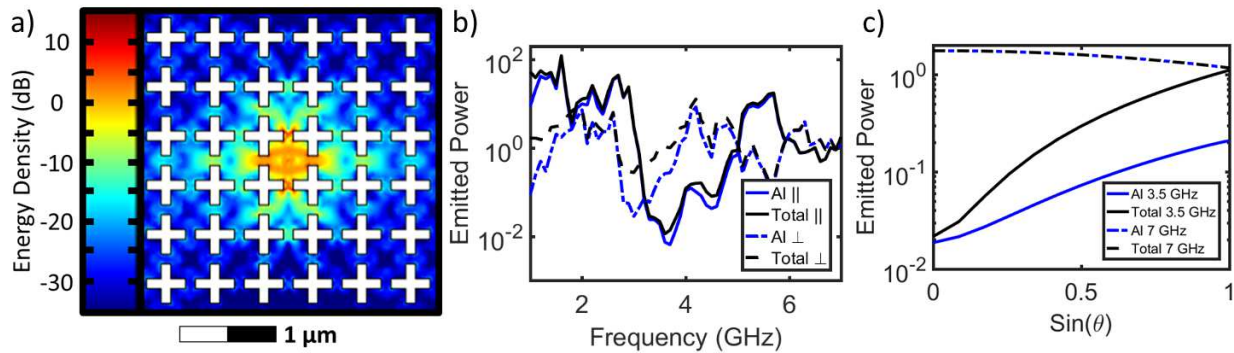}
\caption{(a) Simulation of the phononic bandgap structure at a gapped frequency. The structure is composed of an Al plate connected to a silicon substrate via columns. The simulation is showing the Al plate from above. The white areas are the holes in the Al plate which make up the phononic bandgap structure. The color shows the vibrational energy in the Al due to a point-force in the center. The fast suppression of the energy means energy is prevented from propagating outwards. (b) The power lost by the point-force as a function of frequency, parallel and perpendicular to the Al plane. In blue is plotted the power lost only through the Al layer on top of the substrate. We normalized the results to an unpatterned Al layer directly on top of a Si Substrate. (c) The power radiated as a function of the force’s direction for a frequency inside and outside the bandgap.}
{\label{fig:PB}}
\end{figure*}

In this Letter, we address the phonon bath as an ubiquitous loss channel and model a reduction in qubit coupling to the environment through integrated phononic bandgap structures. Figure \ref{fig:intro} depicts a pedagogical representation for energy transfer between a qubit, a two-level system (TLS) defect,\cite{Phillips1987} and the phonon bath. The electric field from an excited qubit can exchange energy with the TLSs, which are present in amorphous materials and on material interfaces. These TLSs couple to strain fields which then allow them to relax by emitting phonons. The potential value of reducing defect-phonon coupling was proposed in ref. \onlinecite{Agarwal2013}. We expand this insight by developing a comprehensive, multiscale model of the low energy transfer process and identify performance enhancements readily achievable with standard CMOS fabrication techniques. While our simulations are focused on studying the utility of phononic bandgap structures in superconducting resonators and qubits, similar benefits would be expected in quantum dot and dopant-based qubits.

One of the main loss channels for superconducting qubits are two level systems (TLSs).\cite{Martinis2005} The TLS model describes a structural, charge density, or spin reconfiguration inside an amorphous material or on an interface, which can tunnel between two states. When TLSs have a charge reconfiguration, there is a dipole moment difference between the two states, and the TLSs couple to electric fields. Thus, TLSs can exchange photons with energy-degenerate qubits, which are then lost through phonon decay. In the case of continuously driven resonators, the quantized and anharmonic nature of the TLSs is believed to cause the quality factor’s power dependence.\cite{Phillips1987} Due to the quantized nature of the TLSs, and the non-interacting nature of phonons at these wavelengths and single phonon amplitudes, phonon emission must occur at the excited TLS’s resonance frequency, which is the same as the resonator’s frequency for the primary relaxation pathways.

Decreasing the defect mediated vibrational coupling between a qubit and the phonon bath requires a suppression of vibrational modes near the qubit’s operating frequency. To achieve this we evaluated the effects of a phononic bandgap structure. A phononic bandgap is a meta-material composed of repeating unit cells designed to block propagation of certain vibrational modes using Bragg diffraction and Mie resonances.\cite{OlssonIII2009} A wave propagating through the structure at the bandgap frequency will diffract between unit cells and destructively interfere with reflections off the etched areas. The spacing between unit cells is approximately half the wavelength at the bandgap frequency at this condition. This is the ideal frequency structure for quantum information as it blocks resonant phonon emission around the qubit frequency while leaving the low energy density of states unperturbed to maintain thermal conductivity at cryogenic temperatures. Phononic meta-materials with bandgaps in the relevant few GHz regime have previously been demonstrated.\cite{Profunser2009,Safavi-Naeini2014,MayerAlegre2011} It should be noted that if the defects are not true two-level systems but instead anharmonic resonators, they can decay to a lower energy state by emitting a different frequency phonon. However, the broadband phononic bandgaps (several \si{GHz}) will still protect against a large range of possible defect configurations.

In order to demonstrate this method, we modeled a meta-material with a phononic bandgap designed to depress phonon coupling and increase TLS relaxation times (Fig. \ref{fig:PB}). Simulations of the bandgap were performed using the COMSOL Structural Materials Module for a device that suppresses phonon emission over a specific bandwidth. In these simulations we solve the balance of linear momentum equation (Cauchy's first law of motion):

$$\rho\omega^2\vec{u}=\nabla_{x}\sigma+\vec{f_v},$$

where $\omega$ is the solution frequency. $\vec{u}$, the displacement operator, is the measure of the displacement of a point in the material from the undeformed position, and $\rho$ is the deformed material density. $\sigma$ is the Cauchy stress tensor, defined as the force per deformed area in fixed spatial directions, and $\vec{f_v}$ is a body force per unit deformed volume (for more information see Ref. \onlinecite{COMSOL}). Since the materials are all assumed to be linear, we solve the system in the frequency domain. We also assume our materials do not have any dissipation within the simulation cell. Instead, after approximately five wavelengths they enter a Perfectly Matched Layer (PML). This boundary condition models an open system that allows energy to escape the simulation.

The device consists of a \SI{30}{\micro\meter} in radius aluminum (Al) plate patterned with a 2D phononic bandgap. The phononic bandgap structure (Figure \ref{fig:PB}a) was chosen for simplicity of fabrication and analysis and is composed of unit cells of \SI{400}{\nano\meter} by \SI{400}{\nano\meter} Al squares. Each square is connected on four sides to its neighbors via \SI{100x100}{\nano\meter} Al bridges.\cite{Safavi-Naeini2010}Due to simulation size and boundary condition constraints, only the central \SI{5x5}{\micro\meter} are patterned, which corresponds to \num{9x9} unit cells. The Al plate is \SI{100}{\nano\meter} thick and is situated on top of silicon (Si) columns connecting it to a large silicon substrate below. The cylindrical columns are centered under each Al square and have a height of \SI{200}{\nano\meter} and a radius of \SI{100}{\nano\meter}.

A forcing term is placed in the center of the large Al square at the Si column – Al plate interface to represent a TLS. The term is a point source vibrating with a force of \SI{1}{\newton} at the simulation frequency. The power radiating from that forcing term is then integrated at the Al and the Si PML boundaries. The power emitted is proportional to the phonon density of states despite the classical origin of the calculation.\cite{Hwang1999} Figure \ref{fig:PB}b shows the emitted power for the phononic bandgap structure normalized to that of a TLS on an unpatterened Al plate directly in contact with the Si substrate. We study the frequency dependence of two cases – the point force oscillating parallel and perpendicular to the Al plane. In the case of a TLS radiating parallel to the Al plane, we find an order of magnitude reduction in the density of states at \SIrange{3}{5}{\giga\hertz}, the approximate operating frequency of typical qubits. In the case where the dipole radiates perpendicular to the Al plane the bandgap is a lot less effective and shifts to lower frequencies. As expected, the perpendicular TLS radiates power into the Si substrate, whereas in the parallel case most of the power radiates from the Al plate. We also see large variation in perpendicular density as a function of frequency caused by the interference among approximately wavelength-sized structures.

While the electric field is expected to be perpendicular to the conductive superconductor, the complicated amorphous structure of dielectrics won’t necessarily emit phonons in the same direction. Candidate materials for TLSs suggest that in crystalline structures TLS vibrations have a preferential orientation,\cite{Adelstein2017,Holder2013,Gordon2015} but in bulk amorphous materials the orientation will be random. At the surface however, TLS phonon emission may have a preferred direction depending on their microscopic origin. For the purposes of this simulation we assume that the dipole moment is randomly oriented. The projection of the dipole moment in the Z-direction therefore has a distribution that goes as $\sin \theta$, where $\theta$ is zero when the simulation force is parallel to the Al. Figure \ref{fig:PB}c shows the density of states improvement as a function of angle for points inside and outside the bandgap. Performing a weighted average over all the angles gives a decrease in the density of states by \SI{60}{\percent}. 

\begin{figure}
\includegraphics[width=7.5cm]{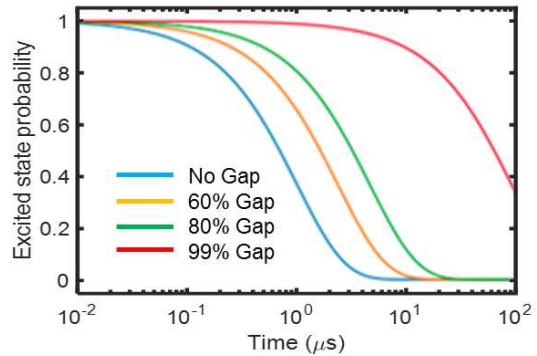}
\caption{The excited state probability of a TLS with intrinsic $T_1=\SI{1}{\micro\second}$ as a function of time under the influence of phononic bandgaps of different size.}
{\label{fig:Model}}
\end{figure}

To calculate the potential improvements in device performance, we consider a simple model of a TLS interacting with the phonon bath. Garraway\cite{Garraway1997} provided a nonperturbative analytic expression for the master equation (ME) of this TLS-boson bath problem in the presence of a bandgap in the bath. Solutions of this ME show improvements in TLS $T_1\approx\frac{1}{1-\Gamma}$, where $\Gamma$ is the density of states suppression due to the bandgap. Figure \ref{fig:Model} shows solutions of this ME for TLSs with an intrinsic (ungapped) lifetime of \SI{1}{\micro\second} coupled to a phonon bath with a \SI{1}{\giga\hertz} bandgap centered on resonance. Here we see that a gap with a \SI{60}{\percent} suppression of the TLS phonon interaction strength extends the lifetime of the TLS by a factor of two, while a \SI{99}{\percent} suppression can extend the lifetime by over two orders of magnitude. In our bandgap structure, assuming a random angle of phonon emission, average $T_1$ increases by a factor of nine and average decay rates, $\Gamma$, decrease by a factor of 2.

This improvement in TLS lifetimes corresponds to an improvement in superconducting circuit performance. The steady state behavior of TLSs on resonators under the influence of a resonant continuous wave electric field is modeled as\cite{Phillips1987}
$$\tan\delta=\frac{\tan\delta_0}{\sqrt{1+\left(\frac{E_{ac}}{E_c}\right)^2}}$$
where $E_{ac}$ is the applied field and the strong-field crossover parameter, $E_c\propto1/\sqrt{T_1T_2}$, is a constant. In steady state, a change in the TLS $T_1$ times will shift the saturation of TLSs to lower applied powers. In other words, in the high power and low temperature (where $T_2=2T_1$) regime the loss tangent will decrease proportionally to the TLS $T_1$ improvement.

We also investigated the transient short time regime. We performed simulations of the open quantum system dynamics with explicit treatment of a finite TLS bath similar to Ref. \onlinecite{Bhattacharya2011}. Using these simulations, we were able to recover short time behavior including coherent backaction between a distribution of TLSs and a single qubit. To prevent the Hilbert space from becoming too large to simulate we initialized the qubit with a single photon and confined our calculation to a subspace with one photon or less. The qubit interacts with \num{200} TLSs through a $\sigma_q^+\sigma_{TLS}^-+\sigma_q^-\sigma_{TLS}^+$ interaction. The TLS simulations include a $\frac{1}{\Delta_0}$ distribution, where $\Delta_0$ is the tunneling energy of a TLS. Theoretically there are an infinite number of TLSs interacting with the qubit with low tunneling energies. However, since the Rabi frequency is proportional to the tunneling energy, those TLSs interact very weakly with the qubit. Assuming $P_0=\SI{5e43}{\per\joule\per\meter\cubed}$,\cite{Khalil2014} a volume of \SI{1e-16}{\meter\cubed} \sisetup{exponent-to-prefix}(\SI{200e-6x200e-6x3e-9}{\meter}\sisetup{exponent-to-prefix=false}), and an energy bandwidth of \SI{10}{\mega\hertz}, \num{200} TLSs allow us to simulate all TLSs with a tunneling energy above $0.01\hbar\omega$. Additionally, as per theory, our TLS $T_1$ times are distributed as $T_1=\frac{T_{1,min}}{\Delta_0^2}$.\cite{Phillips1987} For this simulation we did not assume $T_1$ improvement has an angular dependence.

\begin{figure}
\includegraphics[width=7.5cm]{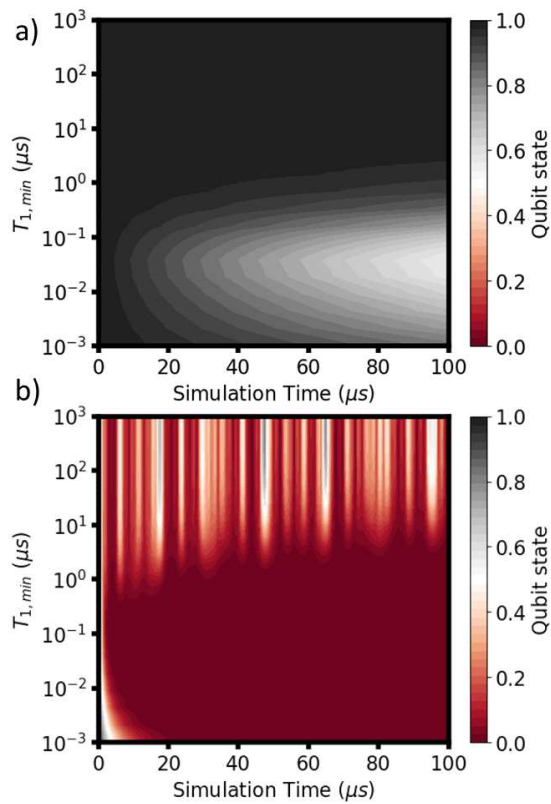}
\caption{Simulations of a qubit initialized with \num{1} photon and propagated in time under the influence of \num{200} TLSs. The x-axis is plotting the simulation time and the color shows the qubit state. $T_{1,min}$, the TLS minimum $T_1$ time, is plotted on the y-axis. The TLSs have the standard distributions in dipole angle and tunneling energy. (a) shows a simulation where TLSs have a maximum Rabi frequency of \SI{45}{\kilo\hertz}. (b) has a maximum Rabi frequency of \SI{450}{\kilo\hertz}.}
{\label{fig:Sim}}
\end{figure}

Electric fields in a typical transmon can vary wildly, from as low as \SI{0.002}{\volt\per\meter} on the top surface of the electrodes to \SI{10}{\volt\per\meter} next to the Josephson junction. Dipole moments are thought to be in the Debye range. While we did not simulate the wide range of possible electric fields, we chose two Rabi frequencies, $\Omega_{rabi,max}=\frac{1}{\hbar}\vec{p}\cdot\vec{E}_{ac}$, that are representative of the observed dynamics. We also included a random dipole orientation which multiplied the chosen Rabi frequency by a random number between zero and one.\cite{Sarabi2016} Figure \ref{fig:Sim} shows the time evolution of the qubit for different $T_{1,min}$, as an increase in this parameter is expected with the phononic bandgap. Figure \ref{fig:Sim}a has a $\Omega_{rabi,max}=\SI{45}{\kilo\hertz}$. At $T_{1,min}\sim\SI{0.1}{\micro\second}$ we see a qubit $T_1$ time on the order of \SI{200}{\micro\second}. As $T_{1,min}$ times are increased, the qubit $T_1$ times increase as well, until at $T_{1,min}\sim\SI{1}{\micro\second}$ we reach a regime where instead of decay, the probability of the qubit staying in its excited state remains above \SI{90}{\percent} for the entire simulation. It is interesting to note that at very low $T_{1,min}<\SI{1}{\nano\second}$, the qubit interaction $T_1$ times increase as well. We believe the simulation is qualitatively correct since with very short $T_{1,min}$, the TLSs approach the over-damped oscillator regime where they are no longer discrete states.\cite{Choi2003}

When we increase $\Omega_{Rabi,max}$ to \SI{450}{\kilo\hertz}, the behavior at $T_{1,min}\sim\SI{0.1}{\micro\second}$ looks like typical qubit decay, but with a $T_1$ time on the order of a microsecond. However, in this regime, increasing the TLS $T_{1,min}$ time has oscillatory behavior. The qubit appears to excite TLSs which then return the energy to the qubit due to the very long decay times. This behavior is observed despite the random distribution in the Rabi frequency. If this behavior is observed in experiments, it could serve as a useful probe for TLS-qubit coupling. There may also be a correction scheme that allows for information recovery since the system doesn’t reach the ground state for a long time. For example, if one were to measure the qubit at \SI{17}{\micro\second}, the qubit state has a \SI{50}{\percent} chance of being excited. There may be a readout scheme that can improve on this by measuring at \SIlist{17;47;65}{\micro\second}. While this spectrum will be system dependent, it may remain stable at low temperatures. However, it is important to recognize that without active fast reset protocols this could lead to long re-initialization times for the system.\cite{Geerlings2013}

To validate the results of the simulation we also derived standard, known, decay rates due to TLSs in bulk. To do this we increased the volume, $V=\SI{6.4e-15}{\meter\cubed}$.\cite{Khalil2014} In order to account for the strongly coupled TLS we increased the simulated number to \num{10000} TLSs. Our Rabi frequency range also changed due to the volume change. For a Rabi frequency of $\Omega_{rabi,max}=\SI{87}{\kilo\hertz}$ ($\SI{870}{\kilo\hertz}$) we measured a $T_1=\SI{580}{\nano\second}$ ($\SI{20}{\nano\second}$) and a quality factor of $Q_i=\num{2900}$ ($\num{100}$). These results are within approximately an order of magnitude of the measured values, and suggest the simulations do reflect the qualitative behavior of the system.

Our simulated design includes protection against TLS decay in the Al layer. We included some 3D protection using columns but very little protection against substrate-based TLSs. This design could be adjusted to reduce loss in tri-layer capacitors. However, for structures such as interdigitated capacitors, where loss is predicted to occur at silicon interfaces,\cite{Wenner2011} 3D protection becomes more involved. Adding phononic protection in between the metal electrodes could further help reduce loss. This could take the form of patterning the Si between electrodes, or using brag mirrors\cite{OlssonIII2009} to protect against emission into the substrate. Alternatively, to protect high-field areas like the Josephson junction in a transmon, the phononically patterned structure can be suspended using an etch process.\cite{Chu2016} However, the ideal case would use a fully 3D bandgap material,\cite{Kitano2015,Johnson2000} which would suppress phonons from TLS emission in any orientation.

In summary, we propose fabrication of superconducting devices with integrated phononic bandgap structures. These structures will increase $T_1$ times of two-level system defects by preventing their relaxation to the environment. We show that this improvement will increase quality factors of superconducting resonators and qubits both in the continuous wave regime and in the single photon limit. In addition to improving quality factors, phononic bandgaps can be used as a probe for the TLSs themselves. Outstanding questions such as whether TLS phonon emission at surfaces have a preferred orientation, or even whether TLS are actual two-level systems as opposed to multi-level anharmonic oscillators could be addressed. Since this technique extends TLS $T_1$ times, it may also be a useful method to access the strong coupling regime\cite{Sarabi2016,Lisenfeld2015,Ramos2013} with TLSs in order to study or control them.

We would like to thank Keith Ray and Robert Sutherland for useful discussions. This work was performed under the auspices of the U.S. Department of Energy by Lawrence Livermore National Laboratory under Contract DE-AC52-07NA27344 and was supported by the LLNL-LDRD Program under Project No. 18-FS-036. LLNL-JRNL-760620

\bibliographystyle{apsrev4-1}
\bibliography{YJR190314_Paper}
\end{document}